\documentclass[pra,aps,twocolumn,superscriptaddress]{revtex4}
\usepackage{epsf}
\usepackage{amsmath}
\usepackage{amssymb}

\newtheorem{lemma}{Lemma}

\newtheorem{criterion}{Criterion}
\newtheorem{proposition}{Proposition}

\newcommand{\beq}{\begin{eqnarray}}
\newcommand{\eeq}{\end{eqnarray}}

\def\<{\langle}
\def\>{\rangle}
\def \ket#1{{| #1 \rangle}}
\def \bra#1{{\langle #1 |}}

\def \tr#1{{\rm Tr}\,\{ #1 \}}
\def \Hat#1{{#1}}

\begin{document}

\title{Inseparability criteria based on matrices of moments}

\author{Adam Miranowicz}

\affiliation{Institute of Theoretical Physics and Astrophysics,
University of Gda\'nsk, 80-952 Gda\'nsk, Poland}

\affiliation{Faculty of Physics, Adam Mickiewicz University,
61-614 Pozna\'n, Poland}

\author{Marco Piani}

\affiliation{Institute of Theoretical Physics and Astrophysics,
University of Gda\'nsk, 80-952 Gda\'nsk, Poland}

\affiliation{Institute for Quantum Computing \& Department of
Physics and Astronomy, University of Waterloo, Waterloo ON,
Canada}

\author{Pawe\l{} Horodecki}

\affiliation{Faculty of Applied Physics and Mathematics, Technical
University of Gda\'nsk, 80-952 Gda\'nsk, Poland}

\affiliation{National Quantum Information Centre of Gda\'nsk,
81-824 Sopot, Poland}

\author{Ryszard Horodecki}

\affiliation{Institute of Theoretical Physics and Astrophysics,
University of Gda\'nsk, 80-952 Gda\'nsk, Poland}

\affiliation{National Quantum Information Centre of Gda\'nsk,
81-824 Sopot, Poland}

\begin{abstract}
Inseparability criteria for continuous and discrete bipartite
quantum states based on moments of annihilation and creation
operators are studied by developing the idea of Shchukin-Vogel
criterion [Phys. Rev. Lett. {\bf 95}, 230502 (2005)].  If a state
is separable, then the corresponding matrix of moments is
separable too. Thus, we derive generalized criteria, based on the
separability properties of the matrix of moments, are thus
derived. In particular, a new criterion based on realignment of
moments in the matrix is proposed as an analogue of the standard
realignment criterion for density matrices. Other inseparability
inequalities are obtained by applying positive maps to the matrix
of moments. Usefulness of the Shchukin-Vogel criterion to describe
bipartite-entanglement of more than two modes is demonstrated: We
obtain some previously known three-mode inseparability criteria
originally derived from the Cauchy-Schwarz inequality, and we
introduce new ones.
\end{abstract}

\date{\today}

\maketitle

\pagenumbering{arabic}
%------------------------------------------------------------------
\section{Introduction}

In recent years, the study of continuous-variable (CV) systems
from the point of view of quantum information has attracted much
interest, stimulated by experimental progress (see
\cite{Braunstein,Braunstein2} and references therein). In
particular, the theory of quantum entanglement for CV systems has
been considerably developed, including the derivation by Shchukin
and Vogel \cite{SV} of a powerful inseparability criterion of
bipartite harmonic quantum states based on Partial Transposition
(PT) \cite{Peres,Horodecki96}, the so-called Positive Partial
Transposition (PPT) criterion. The PPT criterion says that a
separable state remains positive under partial transposition,
therefore a Non-positive-Partial-Transposition (NPT) state must be
entangled. Shchukin and Vogel have demonstrated that their
criterion includes, as special cases, other well-known criteria of
entanglement in two-mode CV systems, including those derived by
Simon \cite{Simon}, Duan {\em et al.} \cite{Duan}, Mancini
\cite{Mancini}, Raymer {\em et al.} \cite{Raymer}, Agarwal and
Biswas \cite{Agarwal}, Hillery and Zubairy \cite{Hillery}. Thus,
the Shchukin-Vogel (SV) criterion can be considered a breakthrough
result, which shows a common basis of many inseparability criteria
for continuous variables (in particular, the results of Duan {\em
et al.} \cite{Duan} seemed previously to be entirely independent
of partial transposition). Another advantage of the SV criterion
should be noted: it is given in terms of creation-operator and
annihilation-operator moments, which are measurable in standard
homodyne correlation experiments \cite{SV2} (for recent reviews on
entanglement detection see Refs. \cite{Horodecki09,Guhne}).

Despite the evident progresses (see also
\cite{Horodecki09,Guhne,pleniocont,others,Horodecki00,Werner} and
references therein), the theory of quantum entanglement for CV
systems can be considered less developed than the theory for
discrete, finite-dimensional systems \cite{Horodecki09}. In the
latter case, powerful inseparability criteria based on positive
maps (see \cite{Horodecki09,Bengtsson} and references therein) and
linear contractions \cite{Rudolph,Chen,Horodecki02,Horodecki03}
(or permutations of the indices of density matrix \cite{Wocjan})
have been studied as generalizations of the standard PPT criterion
\cite{Peres,Horodecki96}. Inspired by these tools available to
study discrete-variable entanglement, we propose a generalization
of the Shchukin-Vogel CV approach.

Shchukin and Vogel \cite{SV} recognized a deep link between the
property of positivity under the operation of PT of a two-mode
density operator $\Hat\rho$, and the positivity under PT of the
corresponding matrix of moments. In the present work, we obtain a
more general relationship between the separability properties of
the density operator and of the matrix of moments. Namely, we show
that if a state is separable, then a suitably designed matrix of
moments is separable too. This will allow us to apply all known
separability criteria (not only the PPT one) to the matrix of
moments rather than directly to the density matrix. For the sake
of clarity, we will analyze explicitly mainly the bipartite
two-mode case; anyway, the results can be extended to the
multimode (see Sect. VII) and multipartite case.

As the objectives of the paper are of wide range, let us first
specify the main goal and results of the paper. We analyze the
Shchukin-Vogel inseparability criterion for matrices of moments
from a new perspective useful for generalizations along the lines
of the standard inseparability criteria for density matrices. More
specifically, we emphasize the fact that separability is preserved
by the mapping from states to matrices of moments. This more
general approach leads us to propose new entanglement criteria
based on realignment and positive maps, which lead to new
inequalities directly applicable in experimental tests of
entanglement.

In particular, in Sect. II, we present a general idea of
separability criteria based on matrices of moments. In Sect. III,
we review the Shchukin-Vogel criterion. In Sects. IV and V, we
present our generalizations of the SV criterion based on the
separability properties of the matrix of moments of creation and
annihilation operators by referring to contraction maps (e.g.,
realignment) and positive maps (e.g., those of Kossakowski, Choi
and Breuer). A few examples illustrating the applicability of the
new criteria are shown. In Sect. VI, we discuss detection of
entanglement by expressing the entries of the density matrix in
terms of the moments. In Sect. VII, we briefly discuss the use of
the criteria to analyze bipartite-entanglement of more than two
modes. Finally, we give our conclusions.

%------------------------------------------------------------------
\section{Separability of states and matrices of moments}
\label{sec:2}

Consider two modes A and B with associated annihilation and
creation operators $a$ and $a^\dagger$ for A and $b$ and
$b^\dagger$ for B. Shchukin and Vogel showed that a Hermitian
operator $X=X^{AB}$ is nonnegative if and only if for any operator
$f=f^{AB}$ whose normally-ordered form exists, i.e., \beq
f=\sum_{k_1,k_2,l_1,l_2=0}^{+\infty}c_{k_1k_2l_1l_2}a^{\dagger k1}
a^{k_2} b^{\dagger l_1} b^{l_2}, \eeq it holds $\langle f^\dagger
f \rangle_X\equiv\tr{f^\dagger f X}\geq 0$.

Let us consider the operators
\begin{eqnarray}
\label{eq:f} \Hat{f}_i\equiv\Hat{f}^A_k\Hat{f}^B_l,
\end{eqnarray}
with $\Hat{f}^A_k\equiv \Hat a^{\dagger k_1} \Hat a^{k_2}$ and
$\Hat{f}^B_l\equiv \Hat b^{\dagger l_1} \Hat b^{l_2}$. Here $i$ is
the unique natural number associated with a double multi-index
$({\bf{k},{l}})$, with ${\bf{k}}=({k}_1,{k}_2)$,
${\bf{l}}=({l}_1,{l}_2)$. Furthermore, the multi-indices
${\bf{k}}$ and ${\bf l}$ are associated with unique natural
numbers $k\leftrightarrow (k_1,k_2)$ and $l\leftrightarrow
(l_1,l_2)$. Any operator $f$ whose normally form exists can thus be written as
$f=\sum_i c_i f_i$. If we further define the matrix $M(\Hat
X)=[M_{ij}(\Hat X)]$, whose elements are given by
\begin{eqnarray}
M_{ij}(\Hat X)\equiv\<\Hat{f}_i^\dagger\Hat{f}_j\>_X=\tr{\Hat
f_i^\dagger \Hat f_j \Hat X}, \label{N01}
\end{eqnarray}
we have
\begin{lemma}
\label{lem:1} An operator $\Hat X$ is positive semidefinite ($\Hat
X\geq0$) if and only if $M(\Hat X)$ is positive semidefinite~\emph{\cite{SV}}.
\end{lemma}
Indeed, $X$ is positive semidefinite if and only if
$\<\Hat{f}^\dagger\Hat{f}\>_X\geq0$ for all $\Hat{f}=\sum_i c_i
\Hat{f}_i$, i.e., if and only if $\sum_{ij} c^*_ic_jM_{ij}(\Hat
X)\geq 0$ for all possible $(c_i)_i=(c_1,c_2,\ldots)$. In turn,
this implies that $X\geq0$ if and only if $M(\Hat X)=[M_{ij}(\Hat
X)]$ is a positive semidefinite (infinite) matrix. We will refer
to correlation matrices as $M(X)$ as to the \emph{matrices of
moments}.

For any density operator $\Hat\rho^{AB}$, from Lemma \ref{lem:1}
we have that the corresponding matrix of moments
$M(\Hat\rho^{AB})$ is positive semidefinite. For a factorized
state $\Hat\rho^{AB}=\Hat\rho^A\otimes\Hat\rho^B$ we have:
\begin{multline}
\label{eq:promoment}
M_{ij}(\Hat\rho^A\otimes\Hat\rho^B)\\
\begin{aligned}
&=\tr{\Hat{f}^\dagger_i\Hat{f}_j\Hat\rho^A\otimes\Hat\rho^B}\\
&=\tr{(\Hat{a}^{\dagger k_{1}}\Hat{a}
^{k_{2}})^\dagger(\Hat{a}^{\dagger
k'_{1}}\Hat{a}^{k'_{2}})(\Hat{b}^{\dagger l_{1}}\Hat{b
}^{l_{2}})^\dagger(\Hat{b}^{\dagger l'_{1}}\Hat{b}^{l'_{2}})
\Hat\rho^A\otimes\Hat\rho^B}\\
&=\tr{(\Hat{a}^{\dagger k_{1}}\Hat{a}
^{k_{2}})^\dagger(\Hat{a}^{\dagger
k'_{1}}\Hat{a}^{k'_{2}})\Hat\rho^A} \tr{(\Hat{b}^{\dagger
l_{1}}\Hat{b }^{l_{2}})^\dagger(\Hat{b}^{\dagger
l'_{1}}\Hat{b}^{l'_{2}})\Hat\rho^B}\\
&=\tr{(\Hat{f}^A_k)^\dagger \Hat{f}^A_{k'}\Hat\rho^A}
\tr{(\Hat{f}^B_l)^\dagger\Hat{f}^B_{l'}\Hat\rho^B}\\
&=M^{A}_{kk'}(\Hat\rho^A)M^{B}_{ll'}(\Hat\rho^B),
\end{aligned}
\end{multline}
where $M^A_{kk'}(\Hat\rho^A)\equiv\tr{(\Hat{f}^A_k)
^\dagger\Hat{f}^A_{k'}\Hat\rho^A}$, so that
$M^A(\Hat\rho^A)=[M^A_{kk'}(\Hat\rho^A)]$ is the matrix of moments
of subsystem A in state $\Hat\rho^A$ (and similarly for B).

A matrix of moments uniquely defines a state, i.e. if
$M(\rho)=M(\sigma)$ then $\rho=\sigma$. This is immediately proven
by considering that if $M(\rho)=M(\sigma)$ then
$\tr{(\rho-\sigma)f^\dagger f}=0$ for all $f$s.

We introduce explicitly formal (infinite) bases~\footnote{We
remark that the introduced bases do not correspond to a Fock
representation, nor are directly related to it.}
$\ket{k}\equiv\ket{{\bf k}}$ and $\ket{l}\equiv\ket{{\bf l}}$, in
which we express the matrices of moments:
\begin{eqnarray}
M(\Hat\rho)=\sum_{kk'll'}M_{kl,k'l'} (\Hat\rho)\ket{k}\bra{k'}
\otimes\ket{l}\bra{l'}. \label{N12}
\end{eqnarray}

Taking into account the one-to-one correspondence between matrices of moments and states and
\eqref{eq:promoment}, we conclude
\begin{proposition}
\label{lem:2} A state is separable, $\Hat\rho=\sum_i p_i
\Hat\rho^A_i\otimes\Hat\rho^B_i$, $p_i\geq0$, $\sum_ip_i=1$, if and only if the
corresponding matrix of moments is also separable, i.e., $M(\Hat\rho)=\sum_i p_i
M^A(\Hat\rho^A_i)\otimes M^B_i(\Hat\rho^A_i)$ with
$M^A(\Hat\rho^A)=\sum_{kk'}M^{A}_{kk'}(\Hat\rho^A)\ket{k}\bra{k'}$
and analogously for $M^B(\Hat\rho^B)$.
\end{proposition}

Notice that the local matrices of moments $M^{A(B)}(\Hat\rho^{A(B)}_i)$ in the Proposition
are physical, i.e., can
consistently be interpreted as related to a local state. Thus, one has to take into
account the subtle point that a matrix of moments could be
separable in terms of generic positive matrices, but not in terms
of physical local matrices of moments. Such a point does not arise when studying the entanglement of
a density matrix: in that case, any convex decomposition in tensor
products of positive matrices is automatically a good physical
separable decomposition. Therefore, it might be that no method
based on the study of separability properties of matrix of
moments, can distinguish all entangled states.

\section{Partial transposition and Shchukin-Vogel criterion}
\label{sec:3}

Let us now recall the Shchukin-Vogel reasoning~\cite{SV}. Let us
first define the operation of partial transposition. Given a
density operator
\begin{eqnarray}
\Hat\rho &=&\sum_{k,l,k',l'}\rho_{k l, k 'l '}|k l \> \< k' l '|
\label{N09}
\end{eqnarray}
in some fixed basis (say in Fock basis), where $\rho_{k l k 'l
'}=\< kl|\Hat\rho|k'l' \>$, its partial transposition (with
respect to subsystem B) is
\begin{eqnarray}
\Hat\rho^\Gamma &=&\sum_{k,l,k',l'} \rho_{k l ,k 'l '}|k l' \> \<
k' l|.\label{N10}
\end{eqnarray}
Partial transposition is a positive but not completely
positive~\footnote{A linear map $\Lambda$ is positive if it preserves
positivity of every matrix on which it acts:
$X\geq0\Rightarrow\Lambda[X]\geq0$. It is moreover completely
positive if it preserves positivity of all matrices on which it
acts partially: $X_{AB}\geq0\Rightarrow({\rm
id}_A\otimes\Lambda_B)[X_{AB}]\geq0$.} linear map which is well
defined also in an infinite-dimensional setting. Positivity of
$\Hat\rho^\Gamma$ is a necessary condition for separability of
$\Hat\rho$~\cite{Peres,Horodecki96}. We rederive explicitly the
relation between the matrix of moments of $\Hat\rho$ and the one
of the partially-transposed state $\Hat\rho^\Gamma$:
\begin{multline}
\label{N04}
M_{kl,k'l'}(\Hat\rho^\Gamma)\\
\begin{aligned}
&=\tr{ (\Hat{a}^{\dagger k_{1}}\Hat{a}
^{k_{2}})^\dagger(\Hat{a}^{\dagger
k'_{1}}\Hat{a}^{k'_{2}})(\Hat{b}^{\dagger l_{1}}\Hat{b
}^{l_{2}})^\dagger(\Hat{b}^{\dagger
l'_{1}}\Hat{b}^{l'_{2}})\Hat\rho^\Gamma}\\
&=\tr{ (\Hat{a}^{\dagger k_{1}}\Hat{a}
^{k_{2}})^\dagger(\Hat{a}^{\dagger
k'_{1}}\Hat{a}^{k'_{2}})\big((\Hat{b}^{\dagger l_{1}}\Hat{b
}^{l_{2}})^\dagger(\Hat{b}^{\dagger
l'_{1}}\Hat{b}^{l'_{2}})\big)^T\Hat \rho}\\
&=\tr{  (\Hat{a}^{\dagger k_{1}}\Hat{a}
^{k_{2}})^\dagger(\Hat{a}^{\dagger
k'_{1}}\Hat{a}^{k'_{2}})(\Hat{b}^{\dagger l'_{1}}\Hat{b
}^{l'_{2}})^\dagger(\Hat{b}^{\dagger
l_{1}}\Hat{b}^{l_{2}})\Hat \rho}\\
&= M_{kl',k'l}(\Hat\rho),
\end{aligned}
\end{multline}
following from the property $\Hat{b}^{T}=\Hat{b}^{\dagger }$.
Therefore, the matrix of moments of the partially-transposed
state corresponds to the partial transpositions of the matrix of
moments of the state. Moreover, considering
Lemma~\ref{lem:1}, we have:
\begin{criterion} ({\bf Shchukin-Vogel}~\emph{\cite{SV}})
A bipartite quantum state $\Hat\rho$ is NPT if and only if
$M(\Hat\rho^\Gamma)=(M(\Hat\rho))^\Gamma$ is NPT. \label{t1}
\end{criterion}
Considering the remarks following Proposition~\ref{lem:2} it is
noteworthy that analyzing the partial transposition of the matrix
of moments we are able to conclude about the PPT/NPT property of
the states. In particular, this means that the only possible
entangled states, for which the analysis of the separability
properties of the corresponding matrix of moments is not enough to
reveal their entanglement, are PPT bound entangled states
\cite{Horodecki97,Horodecki98}.

Given Criterion~\ref{t1}, there is still the problem of analyzing
the positivity of $(M(\Hat\rho))^\Gamma$. Since the matrix of
moments is infinite, one necessarily focuses on submatrices. Let
us define $M_N(\Hat\rho^\Gamma)$ to be the submatrix corresponding
to the first $N$ rows and columns of $M(\Hat\rho^\Gamma)$.
According to the original work by Shchukin and Vogel~\cite{SV}, a
bipartite quantum state would be NPT if and only if there exists
an $N$ such that $\det M_N(\Hat\rho^\Gamma)<0$. As shown in
\cite{MP}, this is not correct, since the sign of all leading
principal minors, i.e., of $\det M_N(\Hat\rho^\Gamma)$, for all
$N\geq1$, does not characterize completely the (semi)positivity of
matrices of moments which are singular. For any (possibly
infinite) matrix $\mathcal{M}$, let $\mathcal{M}_{\bf r}$, ${\bf
r}=(r_1,\ldots,r_N)$ denote the $N \times N$ principal submatrix
which is obtained by deleting all rows and columns except the ones
labelled by $r_1,\ldots,r_N$. By applying Sylvester's criterion
(see, e.g., \cite{Strang}) we find~\cite{MP}:
\begin{criterion}
A bipartite state $\Hat\rho$ is NPT if and only if there exists a
negative principal minor, i.e., $\det (M(\Hat\rho^\Gamma))_{\bf
r}< 0$ for some ${\bf r}\equiv(r_1,\ldots,r_N)$ with $1\le r_1<
r_2<\ldots < r_{N}$. \label{t2}
\end{criterion}

Focusing on the principal submatrix $(M(\Hat \rho))_{\bf r}$, is
equivalent to considering a matrix given by moments $M_{ij}(\Hat
\rho)=\tr{\Hat f^\dagger_i\Hat f_j\Hat\rho}$ only for some
specific operators $\Hat f_i$. In turn, this amounts to study
positivity of $\rho$ (or $\rho^\Gamma$, when we consider $(M(\Hat
\rho^\Gamma))_{\bf r}$) only with respect to a subclass of
operators $f^\dagger f$ (see the proof of Lemma \ref{lem:1}),
i.e., with $f=\sum_{i=1}^Nc_{r_i}f_{r_i}$. Hereafter, if not
otherwise specified, we slightly abuse notation and denote by
${\Hat f}=(\Hat f_{r_1},\Hat f_{r_2},...,\Hat f_{r_N})$ a subclass
of the class of operators \eqref{eq:f}. Let
$M_{f}(\Hat\rho^\Gamma)\equiv(M(\Hat \rho^\Gamma))_{\bf r}$ with
${\Hat f}=(\Hat f_{r_1},\Hat f_{r_2},...,\Hat f_{r_N})$ denote the
principal submatrix corresponding to ${\bf
r}=(r_1,r_2,\ldots,r_N)$. Criterion \ref{t2} can then equivalently
be rewritten as:
\begin{criterion}
A bipartite state $\Hat\rho$ is NPT if and only if there exists ${\Hat f}$
such that $\det M_{f}(\Hat\rho^\Gamma)$ is negative.
\label{t3}
\end{criterion}
More compactly:
\begin{eqnarray}
\Hat \rho \textrm{~is PPT} &\Leftrightarrow& \forall {\Hat f}:
\quad \det M_{ \Hat f}(\Hat\rho^\Gamma) \ge 0,
\nonumber \\
\Hat \rho \textrm{~is NPT} &\Leftrightarrow& \exists {\Hat f}:
\quad \det M_{ \Hat f}(\Hat\rho^\Gamma) <0. \label{N08}
\end{eqnarray}

Notice that in general $M_f(\rho^\Gamma)\neq (M_f(\rho))^\Gamma$,
i.e., the operation of partial transposition and the choice of a
principal submatrix do not commute. The criterion requires to
consider submatrices of the partially-transposed matrix of
moments, i.e., $M_f(\rho^\Gamma)$, not to take submatrices of the
matrix of moments and study their partial transposition.
Nonetheless, also considering the partial transposition of a
submatrix of the matrix of moments is a test for separability, if
the submatrix is chosen in the right way (see Section
\ref{sec:newinsep}, in particular Eq. \eqref{N19a})

On the other hand, for any $f$ (i.e., for any ${\bf r}$), the
moments which constitute the entries of $M_f(\rho^\Gamma)$ and
$M_f(\rho)$, when both expressed with respect to $\rho$, are
simply related by Hermitian conjugation of the mode $b$.

%------------------------------------------------------------------
\section{New inseparability criteria via reordering of matrices of moments}
\label{sec:newinsep}

In this Section, we will be interested in studying the
separability properties of the matrix of moments through a
reordering of its elements. Indeed, apart from partial
transposition, there are other entanglement criteria based on such
reorderings. In the bipartite setting, the only non-trivial one
which is also independent of partial transposition is realignment.
For a state $\Hat\rho$ as in \eqref{N09}, the realigned state
reads
\begin{eqnarray}
\Hat\rho^{R} =\sum_{k,l,k',l'} \Hat\rho_{kl,k'l'}|k k'\> \< l l'
|. \label{N11}
\end{eqnarray}
In a finite-dimensional setting, necessary conditions for
separability can be formulated as $||\Hat\rho^ {\Gamma}||\le 1$
\cite{Peres} and $||\Hat\rho^R||\le 1$ \cite{Rudolph,Chen}, where
$||\Hat A ||=\tr{\sqrt{\Hat A^\dagger \Hat A}}$ is the trace norm
of $\Hat A$. The converse statements, $||\Hat\rho^ {\Gamma}||> 1$
and $||\Hat\rho^R||> 1$, are therefore sufficient conditions for
the state to be entangled. It is worth noting that $||\Hat\rho^
{\Gamma}||\le 1$, contrary to the realignment criterion, is also a
sufficient condition for separability for $2\times 2$ and $2\times
3$ systems \cite{Horodecki96}.

We have seen how the partial transposition of the matrix of
moments corresponds to the matrix of moments of the
partially-transposed state, leading to the SV criterion. It is
immediate to define a realigned matrix of moments following
\eqref{N11}. Unfortunately, there is no simple relation between
the realigned matrix of moments and the realigned state. More
importantly, partial transposition and realignment, while both
corresponding to a reordering of the elements of a matrix, appear
to be on a different footing as regards their applicability in an
infinite-dimensional setting. Indeed, the partial transposition
criterion can be stated as a condition on positivity of the
partially-transposed state/matrix of moments, besides a
condition on the corresponding trace norm. On the other hand, the
realignment condition can be expressed only in the latter way, so
that it is not suited to study the separability properties of a
non-normalized (and non-normalizable) infinite matrix, e.g in the
case of the matrix of moments. To circumvent such an issue, in the
following we will analyze separability properties of properly
truncated matrix of moments, opening the possibility to deploy the
power of the techniques developed for finite-dimensional systems.
We remark that such a ``truncation approach'' could also be
applied directly to CV density matrices, as it was done, for
example, in \cite{Horodecki00}, but in this work we focus on the
matrices of moments. One of the main reasons is that, as already
remarked about SV criterion, moments are measurable in standard
homodyne correlation experiments.

In the SV approach, one typically refers directly to the total
infinite matrix of moments $M(\Hat\rho^\Gamma)$ (see
Criterion~\ref{t1}), studying positivity of its principal minors
(see Criterion~\ref{t2}). Instead, we propose to first truncate
the matrix of moments $M(\rho)$, and then analyze with different
criteria the separability of the truncated matrix of moments.
Indeed, truncation is equivalent to focusing on (some) submatrix.
The submatrix must be chosen correctly, avoiding the introduction
of artifact entanglement by the truncation. The truncated matrix
is positive and, once normalized, can be considered a legitimate
state of an effective bi- or multi-partite finite-dimensional
system. Explicitly, consider subsets of indices
\begin{align*}
I_A&=\{k^{(1)},\ldots,k^{(d_A)}\}\leftrightarrow\{{\bf k}^{(1)},
\ldots,{\bf k}^{(d_A)}\},\\
I_B&=\{l^{(1)},\ldots,l^{(d_B)}\}\leftrightarrow\{{\bf
l}^{(1)},\ldots,{\bf l}^{(d_B)}\}
\end{align*}
and the corresponding projectors $P_A=\sum_{k\in I_A}\ket k \bra
k$ and $P_B=\sum_{l\in I_B}\ket l \bra l$. Then we can define a
finite-dimensional matrix
\begin{eqnarray}
\label{N13} M_{I_AI_B}(\Hat\rho)=\big(P_A\otimes P_B\big)
M(\Hat\rho) \big(P_A\otimes P_B\big)
\end{eqnarray}
and we have that $M_{I_AI_B}(\Hat\rho)/\tr{M_{I_AI_B}(\Hat\rho)}$
is a well-defined state (positive and with trace equal to one) for
a $d_A\otimes d_B$ system, which is separable if the starting
state $\Hat\rho$ is separable. Indeed, according to
Proposition~\ref{lem:2}, if $\Hat\rho$ is separable then
$M(\Hat\rho)$ is separable too; moreover, a further local
projection cannot induce the creation of entanglement.

As we noted at the end of Section~\ref{sec:3}, any choice of a
principal submatrix can be described as considering a specific
class $\Hat{f}$ of operators, i.e., a restricted set of products
of annihilation and creation operators in normal order. Now, we
are interested in the classes of operators corresponding to the
choice of $I_A$ and $I_B$. This means we will always consider only
tensor product classes of operators:
\begin{eqnarray}
\begin{aligned}
{ \tilde f}&=\Hat{f}^A\otimes\Hat{f}^B\\
&=(\Hat{a}^{\dagger k^{(1)}_1}\Hat{a}^{k^{(1)}_2},\ldots,
\Hat{a}^{\dagger k^{(d_A)}_1}\Hat{a}^{k^{(d_A)}_2})\\
&\quad\otimes (\Hat{b}^{\dagger
l^{(1)}_1}\Hat{b}^{l^{(1)}_2},\ldots,
\Hat{b}^{\dagger l^{(d_B)}_1}\Hat{b}^{l^{(d_B)}_2})\\
&=(\Hat{a}^{\dagger k^{(1)}_1}\Hat{a}^{k^{(1)}_2}\Hat{b}^{\dagger
l^{(1)}_1}\Hat{b}^{l^{(1)}_2},\ldots). \label{N14}
\end{aligned}
\end{eqnarray}
With the help of this notation, a truncated matrix of moments will
be denoted in the following as
\begin{eqnarray}
M_{ \tilde f}(\Hat\rho) &\equiv&\sum_{\substack{
k,k'\in I_A\\
l,l'\in I_B}}M_{kl,k'l'}(\Hat\rho)|k l \> \< k' l '| \label{N15}
\end{eqnarray}
for an operator class ${ \tilde f}$, which is given by a tensor
product of classes (as marked by tilde).

Elements of matrix (\ref{N15}) can be reordered to get
entanglement criteria in full analogy to those based on reordering
of the density matrix elements. Thus, we formally apply to $M_{
\tilde f} (\Hat\rho)$ the ``partial transposition''
\begin{eqnarray}
(M_{ \tilde
f}(\Hat\rho))^\Gamma=\sum_{k,l,k',l'} M_{k l k 'l '}(\Hat\rho)|k
'l \> \< k l '|, \label{N16}
\end{eqnarray}
and the ``realignment''
\begin{eqnarray}
(M_{ \tilde f}(\Hat\rho))^R=
\sum_{k,l,k',l'} M_{k l k 'l '}(\Hat\rho) |k k' \> \< l l'
|,\label{N17}
\end{eqnarray}
in complete analogy to (\ref{N10}) and (\ref{N11}). Let us define
the normalized trace norms
\begin{eqnarray}
\nu_{ \tilde f} ^\Gamma (\rho)\equiv  \frac{||(M_{ \tilde
f}(\Hat\rho))^\Gamma||}{\tr{M_{ \tilde f}(\Hat\rho)}},\quad \nu_{
\tilde f} ^{R} (\rho)\equiv \frac{||(M_{ \tilde
f}(\Hat\rho))^R||}{\tr{M_{ \tilde f}(\Hat\rho)}}. \label{N19}
\end{eqnarray}
It is worth noting that, because of the tensor product structure
of $\tilde{f}$, we have
\begin{eqnarray}
  (M_{ \tilde f}(\Hat\rho))^\Gamma=
M_{ \tilde f}(\Hat\rho^\Gamma)
\label{N19a}
\end{eqnarray}
for all $\tilde{f}$ and all $\Hat\rho$.

The SV criterion can now be equivalently formulated as
\begin{criterion}
A bipartite state $\Hat\rho$ is NPT if and only if there exists a
tensor product class $\tilde f$, given by (\ref{N14}), such that
$M_{\tilde f}(\Hat\rho^\Gamma)$ is not positive or, equivalently,
$\nu_{\tilde f} ^\Gamma(\rho)
> 1$.
\label{t4}
\end{criterion}
The Rudolph-Chen-Wu \cite{Rudolph,Chen} realignment criterion for
density matrices, can be generalized straightforwardly for the
matrices of moments as follows:
\begin{criterion}
A bipartite quantum state $\Hat\rho$ is inseparable if there
exists $\tilde f$, such that $(M_{ \tilde f}(\Hat\rho))^R$ has
trace norm $||(M_{ \tilde f}(\Hat\rho))^R||$ greater than
$\tr{M_{\tilde f}(\Hat\rho)}$. \label{t5}
\end{criterion}
More compactly:
\begin{eqnarray}
\Hat \rho \textrm{~is separable} &\Rightarrow& \forall \tilde
f:\quad \nu_{\tilde f} ^{R} (\rho)\le 1,
\nonumber \\
\Hat \rho \textrm{~is inseparable} &\Leftarrow& \exists \tilde f:
\quad \nu_{\tilde f} ^{R}(\rho)
> 1. \label{N21}
\end{eqnarray}
In principle, the criterion \eqref{N21} based on the realignment
of the matrix of moments is inequivalent to the SV criterion based
on PT, similarly as, for finite-dimensional density matrices, the
Peres-Horodecki criterion is not equivalent to the Rudolph-Chen-Wu
criterion.

%------------------------------------------------------------------
\subsection{Exemplary applications of partial transposition and realignment}

Let us give a few examples of application of the inseparability
criteria based on PT and realignment of matrices of moments. We
recall that $(M_{\tilde f}(\rho))^\Gamma=M_{\tilde
f}(\rho^\Gamma)$ for a tensor-product $\tilde f$.

\emph{Example 1.} To detect the entanglement of the singlet state
$|\psi\> =\frac{ 1}{\sqrt{2}}(|01\> -|10\> ),$ one can choose
${\tilde f}=(1,a)\otimes (1,b) \equiv(1,\Hat a,\Hat b,\Hat a\Hat
b)$ yielding the following matrix of moments $M_{\tilde
f}(\Hat\rho)\equiv [M_{ij}]=[ \< \tilde f_i^\dagger \tilde f_j
\>]$:
\begin{equation}
M_{\tilde f}(\Hat\rho)=\left[
\begin{array}{cccc}
1 & \< \Hat a\>  & \< \Hat b\>  & \< \Hat a\Hat b\>  \\
\< \Hat a^{\dag }\>  & \<\Hat  N_a\>  & \< \Hat
a^{\dag}\Hat b\>  & \< \Hat N_a \Hat b\>  \\
\< \Hat b^{\dag }\>  & \< \Hat a\Hat b^{\dag }\>  & \< \Hat
N_b\>  & \< \Hat a\Hat N_b\>  \\
\< \Hat a^{\dag }\Hat b^{\dag }\>  & \< \Hat N_a\Hat b^{\dag }\> &
\< \Hat a^{\dag }\Hat N_b\> & \< \Hat N_a\Hat N_b\>
\end{array} \right] , \label{N35}
\end{equation}
where $\rho=|\psi\>\<\psi|$, and $\Hat N_{a}=\Hat {a}^{\dagger
}\Hat {a}$, $\Hat N_{b}=\Hat {b}^{\dagger }\Hat {b}$ are the
number operators. The only nonzero terms of (\ref{N35}) for the
singlet state are: $M_{11}=1$,
$M_{22}=M_{33}=-M_{23}=-M_{32}=1/2$. Elements of $[M_{ij}]$ can be
reordered, according to (\ref{N16}) and (\ref{N17}), to get
$(M_{\tilde f}(\Hat\rho))^\Gamma$ and $(M_{\tilde f}(\Hat\rho))^R$
equal to
\begin{equation}
\left[
\begin{array}{cccc}
M_{11} & M_{21} & M_{13} & M_{23} \\
M_{12} & M_{22} & M_{14} & M_{24} \\
M_{31} & M_{41} & M_{33} & M_{43} \\
M_{32} & M_{42} & M_{34} & M_{44}
\end{array}
\right] , \left[
\begin{array}{cccc}
M_{11} & M_{12} & M_{21} & M_{22} \\
M_{13} & M_{14} & M_{23} & M_{24} \\
M_{31} & M_{32} & M_{41} & M_{42} \\
M_{33} & M_{34} & M_{43} & M_{44}
\end{array}
\right] ,\label{N36}
\end{equation}
respectively. Thus, for the singlet state one gets the trace
norms, defined by (\ref{N19}), greater than 1, i.e., $\nu_{\tilde
f} ^\Gamma=\nu_{\tilde f} ^{R}=(1+\sqrt{2})/2$, as well as
negative $\det M_{\tilde f}(\Hat\rho^\Gamma)=-1/16$ and $\min {\rm
eig} M_{\tilde f}(\Hat\rho^\Gamma)=(1-\sqrt{2})/2$. It is seen
that both the PT and realignment based criteria detect the
entanglement of the singlet state. It is worth noting that one
could analyze just the submatrix of the first matrix of
(\ref{N36}) corresponding to ${\bf r}=(1,4)$. This amounts to
considering, in the standard SV approach, $M_f(\Hat\rho^\Gamma)$
with ${\Hat f}=(1,\Hat a\Hat b)$. Then one gets
\begin{equation}
M_f(\Hat\rho^\Gamma)\!=\!\left[
\begin{array}{cc}
1 & \< \Hat a\Hat b^{\dag }\>  \\
\< \Hat a^{\dag }\Hat b\>  & \< \Hat N_{a}\Hat N_{b} \>
\end{array} \right], \label{N33}
\end{equation}
from which the Hillery-Zubairy criterion of entanglement follows
\cite{Hillery}:
\begin{eqnarray}
\det M_f(\Hat\rho^\Gamma)=\< \Hat N_{a}\Hat N_{b}\> -|\< \Hat
a\Hat b^{\dag }\> |^{2}\ <0. \label{N34}
\end{eqnarray}
For our state, one gets $M_f(\Hat\rho^\Gamma)=[1,-1/2;-1/2,0]$,
which results in $\det M_f(\Hat\rho^\Gamma)=-1/4$.

\emph{Example 2.} The realignment-based and PT-based criteria can
also detect the entanglement of partially-entangled states. To
show this, let us analyze the state $\ket{\psi}=\frac1{\sqrt{3}}
(\ket{00}+\ket{01}+\ket{10})$ for which negativity is equal to
$2/3$. By choosing $\tilde f$ the same as in Example 1, one gets
\begin{eqnarray}
 M_{\tilde f}(\Hat\rho) = \frac13\left[
\begin{matrix}
     3   &    1  &     1  &     0  \\
     1   &    1  &     1  &     0  \\
     1   &    1  &     1  &     0  \\
     0   &    0  &     0  &     0  \\
\end{matrix}
\right], \label{N40}
\end{eqnarray}
which implies $\nu_{\tilde f}^\Gamma=\nu_{\tilde f}^R=1.1891>1$
(as well as $\det M_{\tilde f}(\rho^\Gamma)=-1/81<0$). Thus, the
entanglement of the state can be detected by both criteria.  As in
Example 1, we can use the submatrix of moments
$M_{f}(\rho^\Gamma)=[1,1/3;1/3,0]$, given by (\ref{N33}) (or,
which is the same, the submatrix $(M_{\tilde f}(\rho^\Gamma))_{\bf
r}$ of the partially-transposed $M_{\tilde f}(\Hat\rho)$ of
(\ref{N40}), for ${\bf r}=(1,4)$), which also has negative
determinant (equal to $-1/9$) and minimum eigenvalue, given by $(3
- \sqrt{13})/6\approx-0.1$.

\emph{Example 3.} The realignment-based criterion is sensitive
also for some infinite-dimensional entangled states, as can be
shown on the example of superpositions of coherent states,
referred to as the two-mode Schr\"odinger cat states,
\begin{eqnarray*}
  |\psi'\>&=& {\cal N}' (|\alpha, -\beta\> - |-\alpha, \beta\>),
\\
 |\psi''\>&=& {\cal N}'' (|\alpha, \beta\> - |-\alpha, -\beta\>),
\end{eqnarray*}
which are normalized by functions ${\cal N}'$ and ${\cal N}''$ of
the complex amplitudes $\alpha$ and $\beta$. As actually shown in
\cite{SV}, the entanglement of $|\psi''\>$ (but also of
$|\psi'\>$) can be detected by the standard SV criterion for $\Hat
f=(1,\Hat b,\Hat a \Hat b)$, for which one gets a negative
determinant $\det M_f(\Hat\rho^\Gamma)$. The realignment-based
criterion applied to the factorized $\tilde f=(1,a)\otimes
(1,b)$ is also sensitive enough to detect entanglement of both
states $|\psi'\>$ and $|\psi''\>$. E.g., for both states with
$\alpha=0.3$ and $\beta=0.2$, one gets the trace norms for
realignment and PT greater than one, i.e., $\nu_{\tilde
f}^R=1.1666$ and $\nu_{\tilde f}^\Gamma=1.1783$. Note again that
by analyzing determinant or minimum eigenvalue of submatrix
$(M_{\tilde f}(\Hat\rho^\Gamma))_{\bf r}$ for ${\bf r}=(1,4)$,
given by (\ref{N33}), one can detect entanglement of the state by
handling less moments.

%------------------------------------------------------------------
\section{Positive maps acting on matrices of moments}

In this section we generalize the SV criterion by applying the
theory of positive maps (see reviews
\cite{Horodecki09,Bengtsson}).

The standard criterion of separability for states which is based
on positive maps says the following~\cite{Peres,Horodecki96}: a
bipartite state $\Hat\rho$ is separable if and only if every
positive linear map $\Lambda$ acting partially (say on the second
subsystem only) transforms $\Hat\rho$ into a new matrix with
nonnegative spectrum, i.e.,
\begin{eqnarray}
  ({\rm id}_A\otimes \Lambda_B )[\Hat\rho^{AB}]\ge 0.
\label{N22}
\end{eqnarray}
(For brevity, the system-identifying superscripts are usually
omitted). Therefore, if the partial action of a positive map on a
state of a composite system spoils the positivity of the state,
then the state must be entangled. Obviously, the Peres-Horodecki PPT
criterion can be formulated as (\ref{N22}), with $\Lambda=T$
being the transposition operation. On the other hand, we note that
realignment is not a positive map, and the related criterion
involves the evaluation of the trace norm of the realigned state,
which is in general not even Hermitian.

One direction of the separability criterion based on positive maps
can be applied in the space of matrices of moments to conclude
that the starting state is entangled. Indeed, the reasoning at the
base of the partial map criterion does not require any
normalization and regards only the property of positivity. More
explicitly:
\begin{criterion}
Let $\Lambda$ be a linear map preserving positivity of (infinite)
matrices, and let $M(\Hat\rho)$ be a separable matrix of moments,
i.e., $M(\Hat\rho)=\sum_n p_n M_n(\Hat\rho^A) \otimes
M_n(\Hat\rho^B)$ with $p_n\geq0$. Then the (infinite) matrix
resulting from the partial action of $\Lambda$, i.e., $({\rm
id}\otimes\Lambda)[M(\Hat\rho)]=\sum_n p_n M_n(\Hat\rho^A) \otimes
\Lambda[M_n(\Hat\rho^B)]$, is also positive. \label{t6}
\end{criterion}
Therefore, if we are given a matrix of moments $M(\Hat\rho)$ for
two modes and a positive map $\Lambda$ and we find that $({\rm
id}\otimes\Lambda)[M(\Hat\rho)]$ is not positive, then we conclude
that the matrix of moments as well as the starting state are not
separable.

If there were a mapping between positive linear maps on states and
positive linear maps on the corresponding matrices of moments, we
could perhaps derive a general theorem of the Shchukin-Vogel type.
Unfortunately such a connection, if existing at all, does not seem
to be immediate. Transposition appears in this sense to be very
special, since transposition of states translates simply into
transposition of matrices of moments. Here, we will limit
ourselves to the application of partial maps to truncated matrices
of moments, so that we have the following:
\begin{criterion}
If, for some $\tilde{f}$, there is a positive linear map $\Lambda$
such that $({\rm id}\otimes\Lambda)[M_{\tilde{f}}(\Hat\rho)]$ is
not positive, then $\Hat\rho$ is entangled.
\end{criterion}

This Criterion is a direct consequence of the observation at the
basis of Proposition 1 and Criterion 6. Essentially, if one
constructs a (sub)matrix of moments that preserves the separable
structure of a state, and finds that the matrix of moments is
entangled (using any arbitrary criterion, in this case linear
maps), then one knows that the state was entangled. We remark that
we are only able to establish a sufficient condition for
entanglement (alternatively, a necessary condition for
separability), contrary to the analogous theorem for density
matrices by Horodecki et al. \cite{Horodecki96}, which says that
there always is a map able to detect the entanglement.

We remark that in the case of transposition, which is defined for
any dimension, the application of the map to the matrix of moments
is equivalent to considering the matrix of moments of the
partially transposed state. Therefore it is possible to directly
focus on submatrices of the form $M_f(\rho^\Gamma)$. On the other
hand, in general, we may consider maps whose action is defined on
finite dimensions: consequently, we have to first take (properly
chosen) submatrices $M_{\tilde{f}}(\rho)$, and only then act
partially on them to obtain $M'_{\tilde{f}}=({\rm id}\otimes
\Lambda)[M_{\tilde{f}}(\Hat\rho)]$. This does not exclude that,
after the action of the map, we may consider the positivity of an
even smaller submatrix $(M'_{\tilde f})_{\bf r}$ of the
partially-transformed submatrix of moments.

For example, one can apply non-decomposable~\footnote{A map is
decomposable if it can be written as
$\Lambda=\Lambda_1^{CP}+\Lambda_2^{CP}\circ T$, where $\circ$
stands for composition, $T$ for the transposition operation and
$\Lambda_i^{CP}$, $i=1,2$ are completely positive maps, which by
definition cannot detect any entangled state if applied partially.
A decomposable map cannot detect the entanglement of a PPT state.}
maps to try to detect the entanglement of PPT entangled states.
Classes of such maps were constructed for arbitrary finite
dimension $N\ge 3$, e.g., by Kossakowski \cite{Kossakowski}, Ha
\cite{Ha}, and recently by Yu and Liu \cite{Yu}, Breuer
\cite{Breuer} and Hall \cite{Hall}.

We are not able to provide examples of PPT bound entangled states,
the entanglement of which is detected by applying positive maps on
submatrices of moments, but the existence of such examples is not
excluded.

Furthermore, we stress that it may happen that a detection method
based on an indecomposable map is able to detect more efficiently
the entanglement of an NPT state than PT itself, e.g. it may be
sufficient to consider smaller submatrices of moments. In any
case, through the application of various indecomposable maps one
can easily generate criteria for separability that are possibly
independent from those obtained from PT. Indeed, as an important
application of the proposed method we stress that it enables a
simple derivation of interesting inseparability inequalities,
e.g., 
%\xxx {\bf I ERASED THE FIRST EQUATION AS IT WAS
%CORRESPONDING TO THE DECOMPOSABLE MAP}
\begin{eqnarray}
2(\<N_a N_b\>+\<N^2_a N_b\>)< |\<N_a b\> - \<a^{\dagger}b \>|^2,
\label{N25b}
\end{eqnarray}
which corresponds to the condition on the determinant of
(\ref{N36a}) obtained in the next subsection.

%------------------------------------------------------------------
\subsection{Exemplary applications of positive maps}

The proposed method can be summarized as follows: First truncate
the matrix of moments, i.e., $M\rightarrow M_{\tilde f}$, then
apply a positive map, i.e., $M_{\tilde f}\rightarrow M'_{\tilde
f}$, and check the positivity of the partially-transformed
submatrix of moments $M'_{\tilde f}$. In turn, this amounts to
considering positivity of submatrices $(M'_{\tilde f})_{\bf r}$,
or, by virtue of Sylvester's criterion, to checking positivity of
determinants ${\rm det}(M'_{\tilde f})_{\bf r}$. Thus, one can say
that submatrices of partially transformed submatrices are
considered.

Here, we give a few examples of application of our inseparability
criteria based on some specific classes of positive maps applied
to matrices of moments.

\subsubsection{Kossakowski and Choi maps}

The Kossakowski class of positive maps transforms matrices
$A=[A_{ij}]_{N\times N}$ in ${\cal C}^N$ onto matrices in the same
space as follows \cite{Kossakowski}
\begin{align}
\Lambda_{K}[A]  = \frac{\openone}{N}{\rm Tr} A+\frac{1}{N-1} g\cdot
(R x+\kappa y{\rm Tr} A), \label{N23}
\end{align}
where `$\cdot$' stands for the scalar product,
$\kappa=\sqrt{(N-1)/N}$, $x=(x_i)_i$, $x_i=\tr{A g_i}$, and
$g=(g_i)_i$ satisfying $g_i=g^*_i$, $\tr{g_i g_j}=\delta_{ij}$,
$\tr{g_i}=0$ for $i,j=1,...,{N^2-1}$. In our applications, we
assume $y=0$, $R$ to be rotations $R(\theta)\in {\rm SO}(N^2-1)$,
and $g_i$ to be generators of SU($N$). Note that the Ha maps
\cite{Ha} do not belong to (\ref{N23}). In a special case for
$A=[A_{ij}]_{3\times 3}$, the Kossakowski map is reduced to the
Choi map \cite{Choi},
\begin{align}
\Lambda_{\rm Choi}[A]  =  -A + \textrm{diag}
 ([ & \alpha A_{11}+\beta A_{22}+\gamma A_{33}, \nonumber \\ &
      \gamma A_{11}+\alpha A_{22}+\beta A_{33}, \nonumber \\ &
      \beta A_{11}+\gamma A_{22}+\alpha A_{33}]), \label{N24}
\end{align}
which is positive if and only if $\alpha \ge 1$, $\alpha +\beta
+\gamma \ge 3$ and $1\le \alpha \le 2 \Rightarrow \beta \gamma \ge
(2-\alpha )^2$, while decomposable if and only if $\alpha\ge 1$
and $1\le \alpha\le 3 \Rightarrow \beta\gamma \ge (3-\alpha)^2/4$.
We denote the resulting (unnormalized) matrix of moments shortly
as
\begin{eqnarray}
  M_{\tilde f}'(\rho)\equiv ({\rm id}\otimes\Lambda_{\rm
Choi})[M_{\tilde f} (\rho)]. \label{N24a}
\end{eqnarray}
It is worth noting that some bound entangled states can be
detected \cite{Horodecki02} by applying to $\Hat\rho$ the
St\"ormer map \cite{Stormer}, which is a special case of the Choi
map for $\alpha=2,\beta=0,\gamma=1$ and of (\ref{N23}) for
$\theta=\pi/3$ and $N=3$.
%\xxx {\bf THE FOLLOWING SENTENCE IS
%COPIED FROM THE OLD EXAMPLE, BUT ACTUALLY IT CAN BE ERASED
%TOGETHER WITH REFS. [41,42]} Choi map with $\alpha=\beta=\gamma=1$
%is decomposable and, actually, corresponds to the reduction map
%$\Lambda[A]=\openone{\rm Tr} A-A$ (see
%Refs.~\cite{reduction_cerf,reduction_horodecki}), thus it is
%weaker than the PT-based criterion.

%\xxx {\bf IT IS A COMPLETELY NEW EXAMPLE. UNFORTUNATELY IT IS MORE
%COMPLICATED THAN THE ONE (IN THE FORMER VERSION) FOR DECOMPOSABLE
%CHOI MAP AND MORE COMPLICATED THAN EQ. (21) BASED ON PT.}

\emph{Example.} As an exemplary application of a positive map, let
us apply the St\"ormer map to $9\times 9$ matrix of moments
$M'_{\tilde f}(\rho)$ for $\tilde f =(1,a,a)\otimes (1,b,b)$. Note
that the chosen map is indecomposable. For simplicity, we analyze
only the submatrix $(M'_{\tilde f}(\rho))_{\bf r}$ for ${\bf
r}=(2,3,7)$:
\begin{eqnarray}
(M'_{\tilde f}(\rho))_{\bf r}= \left[
\begin{matrix}
M_{11}+M_{22}&-M_{23}      &-M_{27}\\
      -M_{32}&M_{22}+M_{33}&-M_{37}\\
      -M_{72}&-M_{73}      &M_{77}+M_{99}\\
\end{matrix}
\right] \nonumber \\
=\left[
\begin{matrix}
1+\<N_a\>          & -\<N_a\>           &-\<a^\dagger b\>\\
-\<N_a\>           & 2\<N_a\>           &-\<a^\dagger b\>\\
-\<a^\dagger b\>^* & -\<a^\dagger b\>^* & \<N_a N_b\>+\<N_b\>\\
\end{matrix}
\right], \label{N40a}
\end{eqnarray}
where  $M_{ij}= \< \tilde f_i^\dagger \tilde f_j \>$ are elements
of the original (not-transformed) matrix of moments,
$M_{\tilde{f}}$. Matrix (\ref{N40a}) for the singlet state is
given by $\frac 12[3,-1,1;-1,2,1;1,1,1]$ having negative
determinant (equal to -1/4), which reveals the entanglement of the
state. Analogously, the entanglement of the partially entangled
state $\ket{\psi}=\frac1{\sqrt{3}} (\ket{00}+\ket{01}+\ket{10})$
can also be detected by (\ref{N40a}), which is now reduced to
$(M'_{\tilde f}(\rho))_{\bf r}=\frac 13[4,-1,-1;-1,2,-1;-1,-1,1]$
with negative determinant (equal to -1/27).

%\xxx {\bf ERASE \emph{Example 2.}}

\subsubsection{Breuer map}

Our inseparability criterion for matrices of moments can also be
based on the Breuer positive map defined in a space of even
dimension $d\ge 4$ as follows \cite{Breuer}:
\begin{eqnarray}
\Lambda_{\rm Breuer}[A]  = {\openone}{\rm Tr} A- A -\vartheta[A],
\label{N24b}
\end{eqnarray}
where $\vartheta[A]=UA^TU^\dagger$ can be interpreted as a time
reversal transformation and is given by a skew-symmetric unitary
matrix $U$. The latter can be constructed explicitly as $U=RDR^T$
in terms of \cite{Hall}:
\begin{equation}
  D = \sum_{k=0}^{d/2-1} e^{i\phi_k}
  (|2k\>\<2k+1|-|2k+1\>\<2k|). \label{N24c}
\end{equation}
for any angles $\phi_k$ and arbitrary orthogonal matrix $R$.
Although antisymmetric unitary matrices exist only in
even-dimensional spaces, the Breuer map can be generalized for
arbitrary dimensions (see, e.g., \cite{Hall}). Thus, it is
tempting to propose an analogous criterion by applying the Breuer
map to a matrix of moments:
\begin{eqnarray}
  M_{\tilde f}''(\rho) &\equiv& ({\rm id}\otimes\Lambda_{\rm
Breuer})[M_{\tilde f} (\rho)] \label{N24d}
\end{eqnarray}
and checking positivity of the transformed matrix $M_{\tilde
f}''(\rho)$. It is worth noting that the Breuer map is a special
case of the Yu-Liu positive map \cite{Yu}, thus even more powerful
and computationally simple inseparability criteria for density
matrices \cite{Yu,Breuer,Hall} can also be applied for matrices of
moments.

\emph{Example 1.} To reveal entanglement of the singlet state, let
us first analyze a matrix $M_{\tilde f}(\Hat\rho)$ of moments
generated by some 16-element $\tilde f$. Antisymmetric unitary
matrix $U$ can, for example, be constructed as the anti-diagonal
matrix
\begin{eqnarray}
 U = \left[
\begin{matrix}
     0  &   0  & 0  & 1  \\
     0  &   0  & 1  & 0  \\
     0  &   -1  & 0  & 0  \\
     -1  &   0  & 0  & 0  \\
\end{matrix}\right].
\label{N24f}
\end{eqnarray}
Then, by applying the corresponding Breuer map,  one can easily
get, from (\ref{N24d}), the transformed $16\times 16$ matrix
$M''_{\tilde f}(\Hat\rho)$ for arbitrary state $\Hat\rho$. This
matrix reveals, for example, entanglement of the singlet state for
various choices of ${\tilde f}$, e.g.: $\tilde f^{(1)}
=(1,a,N_a,a^2)\otimes (1,b,N_b,b^2)$, $\tilde f^{(2)}
=(1,a,N_a,1)\otimes (1,b,N_b,1)$, or even $\tilde f^{(3)}
=(1,a,1,1)\otimes (1,b,1,1)$.

Note that $\tilde f^{(2)}$ and $\tilde f^{(3)}$ do not provide
more information than $(1,a,N_a)\otimes (1,b,N_b)$ and
$(1,a)\otimes (1,b)$, respectively. The matrices of moments
corresponding to the former sets of operators contain redundant
copies of the moments related to the latter sets, i.e., a
repetition of an operator amounts to have a matrix of moments with
repeated columns and rows. We considered such redundant sets of
operators because Breuer criterion requires one of the subsystems
to be at least 4-dimensional, but at the same time we wanted to
emphasize that is possible to detect (by means of Breuer's map)
entanglement with fewer and fewer combinations of ``independent''
operators. We point out that $\tilde f^{(1)}$ provides for sure
more information in general than $\tilde f^{(2)}$, and in turn the
latter more than $\tilde f^{(3)}$.

The entanglement detection can be
much simplified by analyzing the submatrix of $M''_{\tilde
f}(\Hat\rho)$ corresponding, e.g., to ${\bf r}=(2,5)$:
\begin{eqnarray}
(M''_{\tilde f}(\Hat\rho))_{\bf r}= \left[
\begin{matrix}
M_{11}+M_{44}&-M_{25}-M_{47}\\
-M^*_{25}-M^*_{47}&M_{66}+M_{77}\\
\end{matrix}
\right], \hspace{1cm}\label{N24g}
\end{eqnarray}
where, as usual, $M_{ij}= \< \tilde f_i^\dagger \tilde f_j \>$ are
elements of the original matrix $M_{\tilde f}(\Hat\rho)$. For
${\tilde f}={\tilde f^{(1)}}$, matrix (\ref{N24g}) reduces to
\begin{equation}
(M''_{\tilde f^{(1)}}(\Hat\rho))_{\bf r} =\left[
\begin{matrix}
1+\<a^{\dagger 2} a^2\>&-\<a^\dagger b\>-\<a^{\dagger 3} a b\>\\
-\<a^\dagger b\>^*-\<a^{\dagger 3} a b\>^*&\<(1+N_a) N_a N_b\>\\
\end{matrix}
\right].
\end{equation}
For the example of the singlet state, one gets $(M''_{\tilde
f^{(1)}}(\Hat\rho))_{\bf r}=[1,1/2;1/2,0]$, for which the
determinant is $-1/4$. One can get even simpler criterion from
(\ref{N24g}) by choosing ${\tilde f}={\tilde f}^{(2)}$:
\begin{eqnarray}
(M''_{\tilde f^{(2)}}(\Hat\rho))_{\bf r}=\left[
\begin{matrix}
2 & \<N_a b\> - \<a^{\dagger}b \> \\
\<N_a b^\dagger\> - \<a b^{\dagger} \> & \<N_aN_b\>+\<N_a^2N_b\> \\
\end{matrix}
\right]. \label{N36a}
\end{eqnarray}
Explicitly, for the singlet state, we have $\det(M''_{\tilde
f^{(2)}}(\Hat\rho))_{\bf r}=\det [2,1/2;1/2,0]=-1/4$. By contrast to
${\tilde f}^{(1)}$ and ${\tilde f}^{(2)}$, matrix (\ref{N24g}) for
${\tilde f}={\tilde f}^{(3)}$ is positive. Nevertheless entanglement
can be revealed by choosing a larger submatrix of $M''_{\tilde
f^{(3)}}(\Hat\rho)$ corresponding to ${\bf r}=(2,5,7,8)$, which results
in
\begin{eqnarray}
(M''_{\tilde f^{(3)}}(\Hat\rho))_{\bf r}=\left[
\begin{matrix}
2 & x_- & 0 & x_+ \\
x_-^*   & z & y_+^* & 0 \\
0 & y_+ & 2 \<N_b\>& y_- \\
x_+^*   & 0 & y_-^* & z \\
\end{matrix}
\right], \label{N36b}
\end{eqnarray}
where $x_{\pm}=\pm \<b\>-\<a^{\dagger}b \>$, $y_{\pm}=\pm \<a N_b
\> - \<N_b\>$, and $z=\<(N_a+1)N_b\>$. For the singlet state, one
again gets $\det(M''_{\tilde f^{(3)}}(\Hat\rho))_{\bf r}=-1/4$.

It is not surprising that one has to change submatrix (i.e.
\eqref{N36b} instead of \eqref{N24g}), because for $\tilde
f^{(3)}$ less entries of the matrix $M_f(\rho)$ contain
independent information (actually, only a $4\times4$ matrix
(corresponding to $(1,a)\otimes(1,b)$) out of the larger
$16\times16$ matrix (all the other entries are just repetitions)).

\emph{Example 2.} To reveal the entanglement of the Bell state
$\ket{\psi}=\frac1{\sqrt{2}} (\ket{00}+\ket{11})$, one can apply
${\tilde f}={\tilde f}^{(1)}$ or ${\tilde f}^{(2)}$ and the Breuer map to
be the same as in the former example. Here, one can choose
submatrix $(M''_{\tilde f}(\Hat\rho))_{\bf r}$ corresponding to
${\bf r}=(1,6,9)$, which reads as:
\begin{eqnarray}
\left[
\begin{matrix}
M_{2, 2} + M_{3, 3}& -M_{1, 6} - M_{3, 8}& M_{2, 10} + M_{3,
11}\\
-M_{6, 1} - M_{8, 3}& M_{5, 5} + M_{8, 8}& -M_{6, 9} - M_{8,
11}\\
M_{10, 2} + M_{11, 3}& -M_{9, 6} - M_{11, 8}& M_{10, 10} + M_{11,
11}\\
\end{matrix}
\right]. \label{N24h}
\end{eqnarray}
For the analyzed Bell state, (\ref{N24h}) yields $\det(M''_{\tilde
f^{(1)}}(\Hat\rho))_{\bf r}= \det(M{''}_{\tilde f^{(2)}}(\Hat\rho))_{\bf
r}=-1/4$ clearly demonstrating the entanglement.

Thus, it is seen how new inseparability inequalities,
corresponding to $\det (M''_{\tilde f}(\rho))_{\bf r}<0$, can be
obtained by application of positive maps to matrices of moments.

%------------------------------------------------------------------
\section{Detection of bound entanglement of finite-dimensional states through
analysis of moments}

The original SV criterion is based on partial transposition, thus
it cannot reveal PPT bound entanglement. On the other hand, it is
known that the standard realignment criterion applied directly to
the density matrix can detect entanglement of some bound entangled
states \cite{Rudolph,Chen,Horodecki02,Horodecki03,Wocjan}. A
question arises: Can PPT bound entanglement be detected by our
realignment-based generalized criterion? We have tested
numerically some bound entangled states of dimensions $3\times 3$
\cite{Horodecki97,Bennett99}, $2\times 4$ \cite{Horodecki97},
$d\times d$ \cite{Horodecki99,Piani} as well as infinite
\cite{Horodecki00,Werner}, but we have not been able to detect
entanglement by our generalized criterion.

All numerical simulations suggest that the norms of reordered
$M_{\tilde f}$ satisfy the inequality $\nu_{\tilde f}^\Gamma\ge
\nu_{\tilde f}^R$ or, equivalently, $||(M_{\tilde f})^\Gamma||\ge
||(M_{\tilde f})^{R}||$. If this observation is true in general,
then the described realignment-based criterion is useless in
detecting PPT bound entanglement.  Nevertheless, bound
entanglement can be detected via moments with the help of the
formula (see, e.g.,\cite{Wunsche}):
\begin{equation}
\< m_{1}|\Hat\rho |m_{2}\> = {\frac{1}{\sqrt{ m_{1}!m_{2}!}}}
\sum_{j=0}^{\infty }{\frac{(-1)^j}{j!}}\< (\Hat a^{\dag
})^{m_{2}+j}\Hat a^{m_{1}+j}\rangle,  \label{N26}
\end{equation}
which enables calculation of a given density matrix from moments
of creation and annihilation operators. It is worth noting two
properties: (i) The above sum is finite for finite-dimensional
states (ii) Eq. (\ref{N26}) is not convergent for some states of
the radiation field including thermal field with mean photon
number $\geq 1$. The formula readily generalizes for two-mode
fields as
\begin{equation}
\< m_{1},n_{1}|\Hat\rho |m_{2},n_{2}\> =\sum_{j,k=0}^{\infty
}{\frac{ \< (\Hat a^{\dag })^{m_{2}+j}\Hat a^{m_{1}+j}(\Hat
b^{\dag })^{n_{2}+k}\Hat b^{n_{1}+k}\> }{(-1)^{j+k}
j!k!\sqrt{m_{1}!n_{1}!m_{2}!n_{2}!}}}. \label{N27}
\end{equation}
Let us analyze a special case of (\ref{N27}) for two qubits.
Single-qubit annihilation operator is simply the Pauli operator
given by $\Hat a=\sigma ^{-}=[0,1;0,0],$ which implies that there
are only four nonzero terms in sum (\ref{N27}). We can explicitly
write two-qubit density in terms of the moments as follows:
\begin{eqnarray}
\Hat\rho =\left[
\begin{array}{cccc}
\< \overline{N}_{a}\overline{N}_{b}\>  & \< \overline{N} _{a}\Hat
b^{\dag }\> , & \< \Hat a^{\dag }\overline{N}_{b}\> , &
\< \Hat a^{\dag }\Hat b^{\dag }\>  \\
\< \overline{N}_{a}\Hat b\> , & \< \overline{N}_{a}\Hat N_{b}\> ,
& \< \Hat a^{\dag }\Hat b\> , & \< \Hat a^{\dag }\Hat N_{b}\>  \\
\< \Hat a\overline{N}_{b}\> , & \< \Hat a \Hat b^{\dag }\> , &
\< \Hat N_{a}\overline{N}_{b}\> , & \< \Hat N_{a}\Hat b^{\dag }\>  \\
\< \Hat a \Hat b\> , & \< \Hat a\Hat N_{b}\> , & \< \Hat N_{a}\Hat
b\> , & \< \Hat N_{a}\Hat N_{b}\> \end{array} \right], \label{N28}
\end{eqnarray}
where $\overline{N}_{a}=1-\Hat N_{a}$ and $\overline{N}_{b}=1-\Hat
N_{b}$. Matrix (\ref {N28}) can be partially transposed and
realigned. All principal minors of $\Hat\rho ^\Gamma$ are positive
if and only if $\Hat\rho$ is separable. The above simple example
for $2\times 2$ system was given to show the method only. To
detect bound entanglement, one has to analyze at least $2\times 4$
or $3\times 3$ systems. For brevity, we will not present
explicitly density matrices in terms of moments for these systems.
Nevertheless, they can easily be constructed using (\ref{N27}) and
then realigned, according to (\ref{N11}), to detect entanglement
of some bound entangled states \cite{Rudolph,Chen,Horodecki02}.
Finally, let us remark that there are drawbacks of the method: (i)
it works if we know the dimension $d<\infty $ of a given state.
(ii) Usually, it is simpler to directly reconstruct density matrix
rather than to reconstruct it via moments.

%------------------------------------------------------------------
\section{A simple construction of multimode entanglement criteria}

The two-mode SV criterion can readily be applied in the analysis
of bipartite-entanglement of $m$-modes. For this purpose, one can
define an $m$-mode normally-ordered operator
\begin{equation}
\label{eq:fmultimode}
  \Hat f \equiv f(\{\Hat a_i\}) = \sum_{\{n_i\}=0}^{\infty}
  \sum_{\{m_i\}=0}^{\infty} c(\{n_i,m_i\})
  \prod_{i=1}^m (\Hat a_i^{n_{i}})^\dagger \Hat a_i^{m_{i}},
\end{equation}
where for brevity we denote $\{n_i\}\equiv
\{n_1,n_2,\ldots,n_{m}\}$, and similarly other expressions in
curly brackets. As in the proof of Lemma \ref{lem:1}, we have that
an operator $X$ is positive semidefinite if and only if
$\tr{Xf^\dagger f}\geq0$ for every $f$ as in
\eqref{eq:fmultimode}. To analyze how mode $\Hat a_j$ is entangled
to all the other modes, it is enough to identify, in the reasoning
followed in the previous sections, system A with the mode $j$ and
system B with all the other modes. Therefore we take $\Hat a=\Hat
a_j$, while normally-ordered powers $b^{\dagger l_1} b^{l_2}$ are
substituted by normally-ordered powers
\begin{multline*}
\Hat a_1^{\dagger (k_1)_1}\Hat a_1^{(k_1)_2}
\ldots\Hat a_{j-1}^{\dagger (k_{j-1})_1}\Hat a_{j-1}^{(k_{j-1})_2}\\
\Hat a_{j+1}^{\dagger (k_{j+1})_1}\Hat a_{j+1}^{(k_{j+1})_2}
\ldots\Hat a_m^{\dagger (k_m)_1}\Hat a_m^{(k_m)_2}.
\end{multline*}
As in the two-mode setting, we may (and we will) analyze
positivity of an operator $X$ with respect to a restricted class
of operators $f$, more specifically with only some coefficients
$c(\{n_i,m_i\})$ that do not vanish. This corresponds to testing
positivity of principal submatrices.

For example, we show that (\ref{N08}) implies the three-mode
Hillery-Zubairy criterion \cite{Hillery} originally derived from
the Cauchy-Schwarz inequality. By choosing $\Hat f=(1,\Hat a\Hat
b\Hat c)$ (we use the notation introduced in Section \ref{sec:3}),
one gets $M_f(\Hat\rho^\Gamma) = \left[ 1 , \< \Hat a^{\dag }\Hat
b\Hat c\rangle; \< \Hat a\Hat b^{\dag }\Hat c^{\dag }\> , \< \Hat
N_{a}\Hat N_{b}\Hat N_{c}\> \right]$, where $\Hat N_{c}=\Hat
c^{\dag }\Hat c$ and, analogously, $\Hat N_{a}$ and $\Hat N_{b}$
are the number operators. Imposing negativity of the determinant,
one derives
\begin{eqnarray}
\< \Hat N_{a}\Hat N_{b}\Hat N_{c}\> < |\< a^{\dag }\Hat b\Hat c\>
|^{2}, \label{N31}
\end{eqnarray}
which is the desired Hillery-Zubairy criterion \cite{Hillery},
i.e., a sufficient condition for the state to be entangled. By
restricting the above case to two modes (say $\Hat c=\Hat 1$), one
can choose $\Hat f=(1,\Hat a\Hat b)$, which leads the
Hillery-Zubairy two-mode entanglement condition \cite{Hillery},
given by (\ref{N34}), as already shown in \cite{SV}. By choosing a
different function $\Hat f$, one can obtain new
Hillery-Zubairy-type three-mode criteria. For example, let us
choose $\Hat f=(\Hat a,\Hat b\Hat c)$ then $M_f(\Hat\rho^\Gamma)=[
\< \Hat N_{a}\> , \< \Hat a\Hat b\Hat c\> ; \< \Hat a\Hat b\Hat
c\> ^{\ast } , \< \Hat N_{b}\Hat N_{c}\rangle]$, which results in
a sufficient condition for the three-mode entanglement:
\begin{eqnarray}
\< \Hat N_{a}\> \< \Hat N_{b}\Hat N_{c}\> < |\< \Hat a\Hat b\Hat
c\> |^{2}. \label{N32}
\end{eqnarray}
In a special case, (\ref{N32}) is reduced to another two-mode
entanglement condition of Hillery and Zubairy: $\< \Hat N_{a}\> \<
\Hat N_{b}\> < |\< \Hat a\Hat b\> |^{2}$, derived from the
Cauchy-Schwarz inequality in \cite{Hillery}.

%------------------------------------------------------------------
\section{Conclusions}

We have studied inseparability criteria for  bipartite quantum
states, which are given in terms of the matrices of observable moments of
creation and annihilation operators, therefore generalizing the analysis by
Shchukin and Vogel.
Indeed, we have suggested (also by means of examples) that all
the techniques
originally developed to detect ``directly''---that is, by considering the
physical density matrix---the entanglement of states, can
be deployed at the level of the matrices of moments. In doing this there
are advantages---e.g., by considering an appropriate submatrix of the
matrix of moments one can apply techniques developed for finite
dimensional system to detect the entanglement of infinite-dimensional
systems---and disadvantages---e.g., while the separable structure
of an entangled state is inherited by all properly constructed matrices
of moments, it is not completely clear how the entanglement of the
starting physical state gets encoded in the matrix of moments, and
in some cases it may be difficult to choose the correct technique to detect it.

In particular, we have proposed a new criterion based on
realignment of elements of the moment matrices of special symmetry
(i.e., corresponding to tensor product $\tilde f$s), as a
generalization of the Rudolph-Chen-Wu realignment criterion
applied for density matrices. Another reordering of elements of
the moment matrices corresponds to the partial transposition as in
the original SV criterion. We have proposed another criterion
based on positive maps applied to appropriate submatrices of
moments. We further observe that the formalism of matrices of moments
can be certainly
combined with the powerful criterion invented in the finite-dimensional
setting by Doherty~{\em et al.}~\cite{Doherty}, in the attempt to
 detect, e.g., the entanglement of continuous-variable systems. How
 powerful this combination can be is nonetheless not evident
 or easily predictable, and we leave it as an interesting open problem.

We have also discussed applications of the SV criteria to describe
bipartite-entanglement of more than two modes. In particular, we
have obtained the three-mode Hillery-Zubairy criteria originally
derived from the Cauchy-Schwarz inequality, and derived new ones
of the same type.

As regards the confidence in the certification of entanglement,
if entanglement is verified within error
bars for the matrix of moments (e.g., by considering the determinants of
submatrices of the partially-transposed matrix of moments as in the
original SV criterion), then
entanglement is certified for the physical state. This is true both
in the case where error bars come from uncertainties in an
experiment---from which the entries of the matrix of moments are
obtained---or from numerical tools. We remark that here we
are just considering certification of entanglement: in this paper we
have not
explored the relation between the degree of entanglement---as
quantified by some entanglement measure---of the physical state
and the degree of entanglement of the matrix of moments.

In conclusion, although it is an open question whether our criteria generalizing
the Shchukin-Vogel idea are sensitive enough to detect bound
entanglement, they enable to derive new classes of classical
inequalities, which can be used for practical detection of quantum
entanglement.

\emph{Note added}. After completion of the first version of our
paper, the SV criterion was thoroughly applied to the multipartite
CV case in~\cite{multiSV}.

%------------------------------------------------------------------
{\bf Acknowledgments}. We would like to thank Micha\l{} Horodecki
for useful comments and observations, and for reading an early
version of the manuscript. We also thank  Jens Eisert, Otfried
G\"uhne, Karol Horodecki, Maciej Lewenstein, Adam Majewski, Anna
Sanpera and Karol \.Zyczkowski for comments. This work was
supported by grant PBZ MIN 008/P03/2003, EU grants RESQ (IST 2001
37559), QUPRODIS (IST 2001 38877) and EC IP SCALA. MP was also
supported by NATO-CNR Advanced Fellowship.

\end{document}